\documentclass[conference,10pt]{IEEEtran}
\usepackage{booktabs}
\usepackage{graphicx}
\usepackage{subfig}
\usepackage{amsmath}
\usepackage{amsfonts}
\usepackage{color}
\usepackage{geometry}
\usepackage{epstopdf}
\DeclareRobustCommand*{\IEEEauthorrefmark}[1]{\raisebox{0pt}[0pt][0pt]{\textsuperscript{\footnotesize #1}}}
\begin{document}
%
\title{Spatial Separation of Closely-Spaced Users in Measured Distributed Massive MIMO Channels}

\author{\IEEEauthorblockN{
Yingjie Xu\IEEEauthorrefmark{},
Michiel Sandra\IEEEauthorrefmark{},
Xuesong Cai\IEEEauthorrefmark{}, 
Sara Willhammar\IEEEauthorrefmark{},   
Fredrik Tufvesson\IEEEauthorrefmark{}      
}                                     
\IEEEauthorblockA{\IEEEauthorrefmark{}
Department of Electrical and Information Technology, Lund University, Lund, Sweden.\\
E-mail: \{yingjie.xu, michiel.sandra, xuesong.cai, sara.willhammar, fredrik.tufvesson\}@eit.lth.se}
}



\maketitle

\begin{abstract}
Aiming for the sixth generation (6G) wireless communications, distributed massive multiple-input multiple-output (MIMO) systems hold significant potential for spatial multiplexing. In order to evaluate the ability of a distributed massive MIMO system to spatially separate closely spaced users, this paper presents an indoor channel measurement campaign. The measurements are carried out at a carrier frequency of 5.6~GHz with a bandwidth of 400~MHz, employing distributed antenna arrays with a total of 128 elements. Multiple scalar metrics are selected to evaluate spatial separability in line-of-sight, non line-of-sight, and mixed conditions. Firstly, through studying the singular value spread, it is shown that in line-of-sight conditions, better user orthogonality is achieved with a distributed MIMO setup compared to a co-located MIMO array. Furthermore, the dirty-paper coding (DPC) capacity and zero forcing (ZF) precoding sum-rate capacities are investigated across varying numbers of antennas and their topologies. The results show that in all three conditions, the less complex ZF precoder can be applied in distributed massive MIMO systems while still achieving a large fraction of the DPC capacity. Additionally, in line-of-sight conditions, both sum-rate capacities and user fairness benefit from more antennas and a more distributed antenna topology. However, in the given NLoS condition, the improvement in spatial separability through distributed antenna topologies is limited. 

\end{abstract}

{\smallskip \keywords 6G wireless communication, distributed massive MIMO, multi-user MIMO, channel measurements, user spatial separability}

%
\IEEEpeerreviewmaketitle

\section{Introduction}
As a new paradigm in multiple-input multiple-output (MIMO) systems, distributed massive MIMO system deploys antenna elements over a large geographical area~\cite{U. Madhow2014}. When integrated with the architecture of cell-free networks, it is expected to further mitigate the degradation of the performance of cell-edge users and support even better uniform coverage~\cite{H. Q. Ngo2017}. 

In a typical multi-user MIMO (MU-MIMO) system, the user spatial separability, which signifies the capability to serve many users simultaneously, is closely associated with the spatial degrees of freedom (DoF) in the propagation channels. Studies have theoretically found that the spatial DoF can be improved by increasing the antennas at the base station (BS)~\cite{E. G. Larsson2014,F. Rusek2013, E. Bjornson2019}. With a large number of antennas, the channels between different users tend to be orthogonal, leading to pair-wise favorable propagation~\cite{F. Rusek2013}. This reduces intra-cell interference between simultaneous users, resulting in improved user spatial separability. This characteristic was investigated in~\cite{X. Gao2015} through practical massive MIMO channel measurements and compared to conventional 8$\times$8 MU-MIMO channels. Experiments in~\cite{J. Flordelis2018} further concluded that even in the challenging case of LoS condition, already 18~closely-located users can be served concurrently by a BS equipped with 128~antennas, demonstrating notable user spatial separability. 

The aforementioned studies were mainly based on co-located massive MIMO architectures. Previous research in~\cite{U. Madhow2014, O. T. Demir2021} claimed that since channels between distributed BSs or access points (APs) are less correlated, distributed massive MIMO facilitates the exploitation of favorable propagation. The degree of favorable propagation was analyzed in~\cite{Z. Chen2018} for a distributed massive MIMO system, where the effects of distances between users and antenna topologies on spatial separability were further discussed. 
In \cite{V. Croisfelt2022}, the conditions if a user can be viewed as spatially separable were investigated, and random access protocols were proposed, aided by the spatial separability of distributed massive MIMO. Note that the studies in~\cite{Z. Chen2018, V. Croisfelt2022} are based on synthetic channels. The achievable spectral efficiency of separable users was discussed in~\cite{T. Choi2020} through measured channels. However, in this study, only line-of-sight (LoS) propagation was considered, and the users in the measurement were widely distributed, ignoring spatial separation in more challenging cases, such as closely spaced users.

To the best of our knowledge, a comprehensive investigation of the spatial separation of closely spaced users based on measured distributed massive MIMO channels is still missing in the literature. To fill this gap, this paper first presents an indoor distributed massive MIMO channel measurement campaign. Then singular value spread (SVS), dirty paper coding (DPC) capacity, and zero forcing (ZF) sum-rate capacities are introduced as metrics to assess spatial separability. The effects of the number of antennas, antenna topologies, and precoding schemes on user spatial separability are examined.

The remainder of this paper is organized as follows. The distributed massive MIMO measurement campaign is described in Section~II. Then in Section~III, the signal model and spatial separation metrics are introduced. Section~IV outlines the evaluation of user separability based on the measurement data. Finally, conclusions are drawn in Section~V.

\section{Measurement Environment and Setup }
The measurements were made with a carrier frequency of 5.6~GHz and a bandwidth of 400~MHz. A wideband USRP-based distributed massive MIMO channel sounder was employed, consisting of NI USRP X410, SP16T RF switches, Rubidium clocks, host computers, and antennas~\cite{M. Sandra2024}. The user equipment (UE) was configured with a monopole antenna mounted on a 1.2~m-height robot that can move along arbitrary trajectories. Zadoff-Chu sequence was selected as the sounding signal with a length of 1024. At the BS end, a total of eight uniform planar arrays, here denoted as panels, were utilized. Each panel was equipped with $2\times4$ dual-polarized patch elements (16~ports in total). Rubidium clocks were applied to synchronize the USRPs at the BS and UE. Before measurements were taken, back-to-back calibration was performed to eliminate responses of the measurement system, connectors, and cables. 

\begin{figure}[tb]
	\centering
	\subfloat[]
	{
		\begin{minipage}[tb]{0.23\textwidth}
			\centering
			\includegraphics[width=1\textwidth]{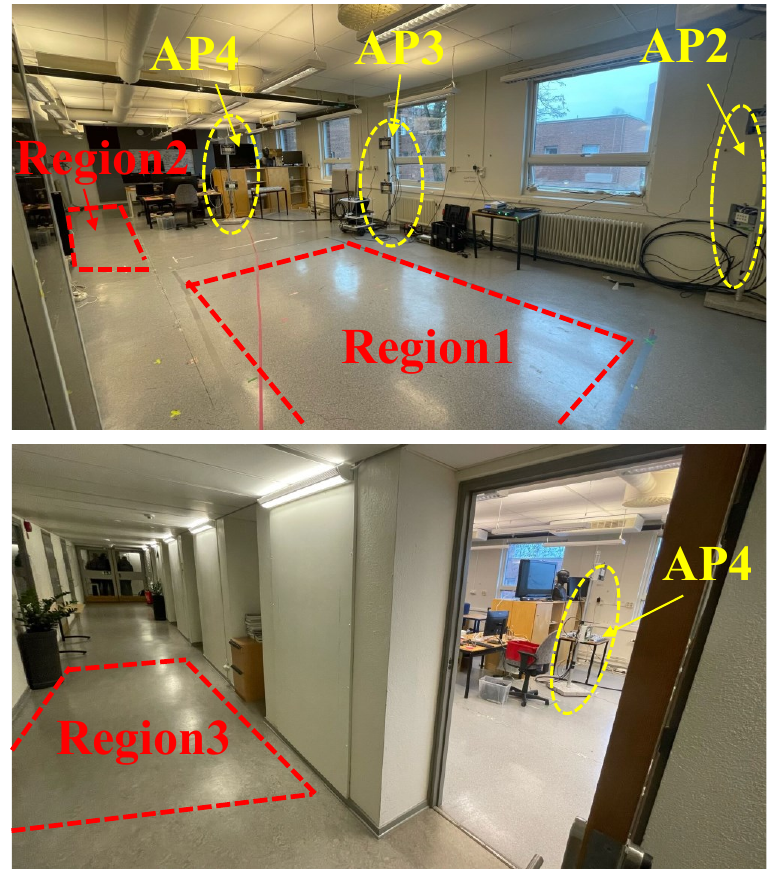}
			\label{Fig_Photo_of_Environments}
		\end{minipage}
	}
	\subfloat[]
	{
		\begin{minipage}[tb]{0.22\textwidth}
			\centering
			\includegraphics[width=1\textwidth]{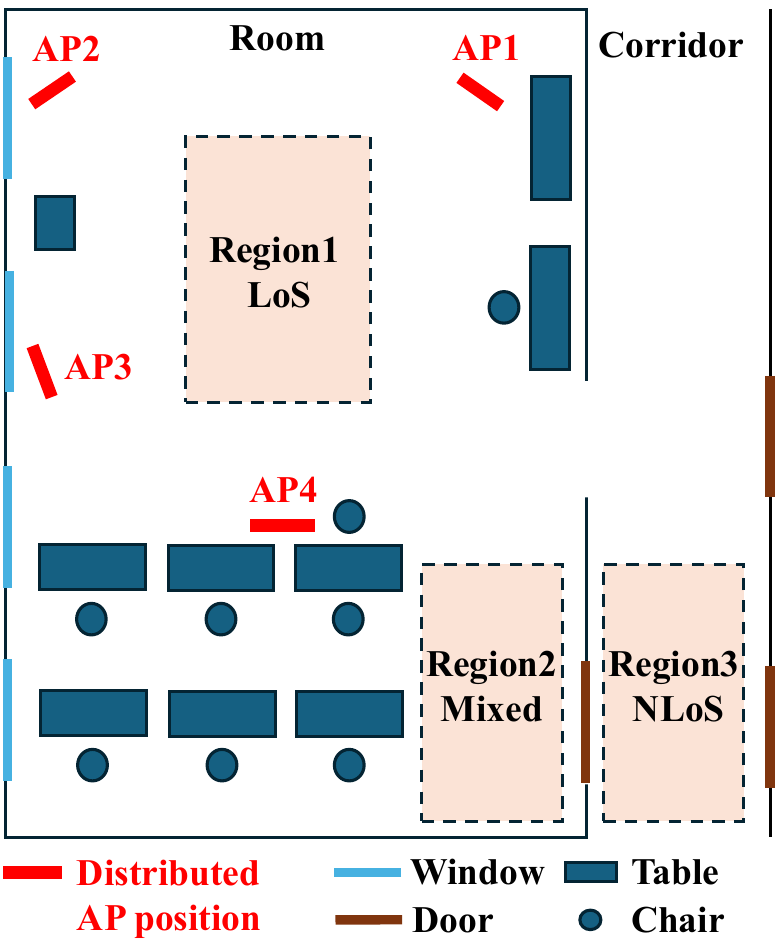}
			\label{Fig_Layout_of_Environments}
		\end{minipage}
	}
	\caption{(a) Photos of the measurement environments and (b) a schematic of the measurement regions.}
	\label{Fig_Photo and layout of the measurement}
\end{figure}

Channel data was collected in an office room and in its adjacent corridor area, as depicted in Fig.~\ref{Fig_Photo and layout of the measurement}(a). On the Rx side, eight panels were divided into four groups that were distributed and each group was viewed as an AP. In this way, each AP consists of 32 antenna elements. They were fixed in the room as shown in Fig.~\ref{Fig_Photo and layout of the measurement}. In this work, three small regions were measured: the front and side areas of the room and the corridor area, as illustrated in Fig.~\ref{Fig_Photo and layout of the measurement}(b). Region1 and Region3 exhibited LoS and non-LoS (NLoS) conditions for all distributed APs, respectively. Region2 showed LoS propagation to the APs located at AP1 and AP2 while manifesting NLoS propagation to the other APs. For simplicity, they are named sequentially as the LoS, mixed, and NLoS regions. Equipped with a single transmit antenna, the robot moved randomly within the measurement regions. Positions from the robot's trajectory were selected to mimic `virtual' closely spaced users. Detailed parameter settings for the channel sounder and measurement configurations are listed in Table~\ref{Measurement Parameters}.      
\begin{table}
\centering
\caption{Parameter settings for the channel measurements.}
\label{Measurement Parameters}
\begin{tabular}{cc}
\toprule 
Parameter & Values   \\
\midrule
Center frequency         & 5.6 GHz       \\
Bandwidth         & 400 MHz         \\
Sounding waveform length          & 2 ns        \\
Delay resolution & 2.5 ns \\
Snapshot rate & 20 Hz \\
Number of Tx (Rx) antennas & 1 (128) \\
& LoS region: 2.5$\times$5 $\rm{m}^2$ \\
Measurement area&Mixed region: 1.7$\times$5 $\rm{m}^2$\\
&NLoS region: 2$\times$5 $\rm{m}^2$\\
\bottomrule
\end{tabular}	
\end{table}

\addtolength{\topmargin}{0.05in}
\section{Signal Model and Metrics of Spatial User Separability}
Consider a multi-user distributed MIMO system with a total number of antennas $M$, where $N$ distributed APs aim to serve $K$ single-antenna users. Define $\mathbf{s}_{t,l}$ as the transmit signal vector in subcarrier $l$ and snapshot $t$. In this paper, the array gain is harvested as a reduced transmit power, which is achieved by applying the transmit power constraint $\mathbf{E}\left\lbrace\mathbf{s}_{t,l}^{\text{H}}\mathbf{s}_{t,l}\right\rbrace=\frac{\rho K}{M}$, where $\rho$ is the mean signal-to-noise ratio (SNR) at the user side. Then the $K\times1$ received signal model can be written~as 
\begin{equation}
\begin{split}
\mathbf{y}_{t,l}=\mathbf{H}_{t,l}\mathbf{s}_{t,l}+\mathbf{w}_{t,l}
\end{split}
\label{Signal model}
\end{equation}
where $\mathbf{H}_{t,l}=\left\lbrack \mathbf{h}_{t,l,1},\mathbf{h}_{t,l,2},...,\mathbf{h}_{t,l,K}\right\rbrack^{\text{T}}\in{\mathbb{C}^{K\times M}}$ represents the channel matrix, and $\mathbf{w}_{t,l}$ is assumed to be independent and identically distributed (i.i.d.) complex white Gaussian noise with unit variance. Note that the channel matrix will be normalized after being obtained through measurement. The $k$th measured user channel vector $\mathbf{h}_{t,l,k}$ is normalized through~\cite{X. Gao2015}
\begin{equation}
\begin{split}
\mathbf{h}_{t,l,k}^{\mathrm{norm}}=\sqrt{\frac{128LT}{\displaystyle{\sum_{l=1}^{L}\sum_{t=1}^{T}||\mathbf{h}_{t,l,k}^{\mathrm{meas}}||_{F}^{2}}}}\mathbf{h}_{t,l,k}^{\mathrm{meas}}, k=1,2,...,K.
\end{split}
\label{Normalization}
\end{equation}
The normalization is performed to remove the imbalance of channel attenuations between different users while preserving the fading impact over the subcarriers and symbols. Note that the variations in channel gain over the distributed antenna elements are also preserved. These variations, caused by large-scale fading/antenna orientation/relative position, are critical for the performance evaluation of distributed massive MIMO. 
\subsection{Singular Value Spread}
As mentioned in Section~I, the user separability relies on favorable propagation, where user channels tend to be orthogonal. To evaluate to what degree real distributed MIMO multi-user channels are “favorable'', the SVS of the channels is investigated. It reflects the joint orthogonality of all users, and can be calculated using the singular value decomposition (SVD) of the normalized channel matrix, that is, $\mathbf{H}_{t,l}^{\mathrm{norm}}=\mathbf{U}_{t,l}\mathbf{\Sigma}_{t,l}\mathbf{V}_{t,l}^{\mathrm{H}}$,
where $\mathbf{U}_{t,l}\in{\mathbb{C}^{K\times K}}$ and $\mathbf{V}_{t,l}\in{\mathbb{C}^{M\times M}}$ denote the unitary matrices and $\mathbf{\Sigma}_{t,l}=\text{diag}(\sigma_{1},...\sigma_{K})$ is the diagonal matrix containing the singular values of $\mathbf{H}_{t,l}^{\mathrm{norm}}$. Then the SVS in logarithmic units can be expressed as 
\begin{equation}
\begin{split}
\kappa_{t,l}=10\text{log}_{10}{\frac{\operatorname*{max}_{k}\sigma_{k}}{\operatorname*{min}_{k}\sigma_{k}}}.
\end{split}
\label{SVS}
\end{equation}
A larger SVS means a stronger linear dependency between at least two rows of $\mathbf{H}_{t,l}$, meaning that there are at least two user channels that exhibit stronger correlations. When $\kappa_{t,l}=0$, optimal orthogonality among all users is achieved, resulting in favorable propagation and contributing to good user separability of co-scheduled users.  

\subsection{DPC Capacity}
To comprehensively quantify the user separability in the distributed MIMO system, the multi-user sum-rate capacity is selected as a second indicator. It is achieved by dirty-paper coding (DPC)~\cite{M. Costa1983}. Assuming perfect channel state information (CSI) at both Tx and Rx, the DPC capacity is given by~\cite{J. Flordelis2018}
\begin{equation}
\begin{split}
C_{DPC}=\operatorname*{max}_{p_{k}}\frac{1}{LT}\sum_{
l=1}^{L}\sum_{t=1}^{T}\log_{2}\mathrm{det}(\mathbf{I}_{M}+\frac{\rho K}{M}\mathbf{H}_{t,l}^{\text{H}}\mathbf{PH}_{t,l})
\end{split}
\label{DPC_sumrate capacity}
\end{equation}
where $\mathbf{P}=\text{diag}(p_{1},...p_{K})$ is a power allocating matrix with total power constraint $\operatorname{tr}(\mathbf{P})=1$. The optimal capacity and user power allocation in (\ref{DPC_sumrate capacity}) can be found through sum-power iterative water-filling algorithms~\cite{N. Jindal2005}.  
\subsection{ZF Linear Precoding Sum-Rate Capacity}
Although the DPC precoder can achieve the multi-user sum-rate capacity, it shows significantly high computational complexity, especially when a massive number of antennas are deployed in the system. Considering practical deployments, the low-complex zero forcing (ZF) linear precoding scheme has been shown to efficiently eliminate user interference~\cite{T. Yoo2005}. It is selected here as the third metric for user separability. The sum-rate capacity with the ZF precoding scheme can be obtained as the solution to the following convex problem through a standard water-filling algorithm~\cite{J. Flordelis2018}.    
\begin{gather}
   C_{ZF}=\operatorname*{max}_{p_{k}}\frac{1}{LT}\sum_{n=1}^{N}\sum_{t=1}^{T}\sum_{k=1}^{K}\log_{2}(1+\frac{\rho K}{M}\frac{p_{k}}{g_{k,t,l}^{2}})
\label{ZF_sumrate capacity}
\end{gather}
where $\sum_{k=1}^{K}p_{k}=1$, and $g_{k,t,l}^{2}$ is the $(k,k)^{\text{th}}$ element of matrix $(\mathbf{H}_{t,l}\mathbf{H}_{t,l}^{\text{H}})^{-1}$.  

\section{Results and Discussion}
\begin{figure}[tb]
	\centerline{\includegraphics[width=0.48\textwidth]{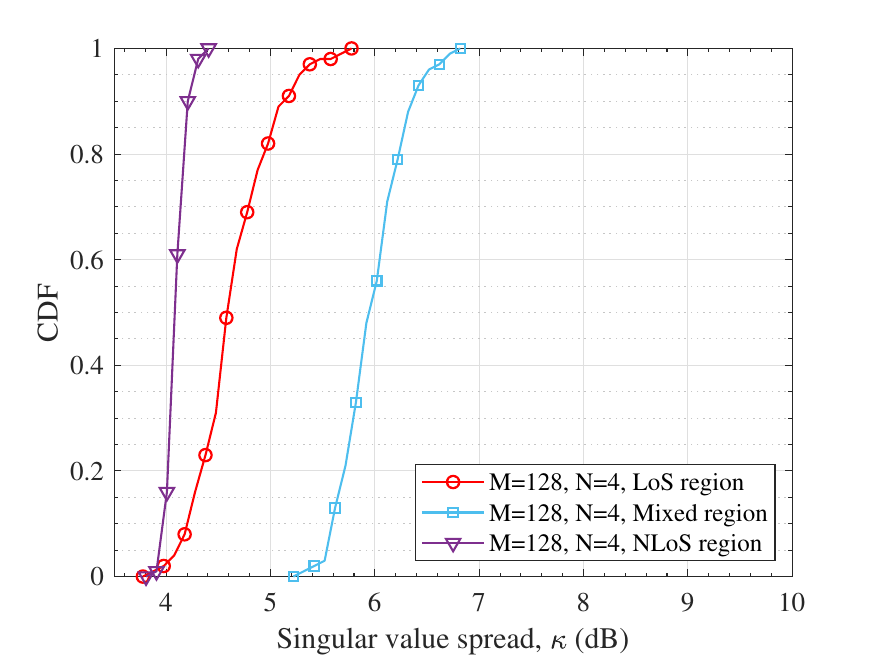}}
	\caption{CDFs of SVSs measured in different regions (${M} = 128$, ${N} = 4$, $K=12$).}
	\label{Fig_SVS_DifferentRegion_12user}
\end{figure}

In this section, the spatial separability of the system is evaluated using the metrics introduced in Secion~III. To study the effects of the number of antennas and their topologies on spatial separability, channel data are chosen from subarrays of different APs. The selection criteria are as follows. For a system with a total number of antennas $1\le M\le128$, the APs are first selected from the $N=1,2,3$ and $4$ locations shown in Fig.~\ref{Fig_Photo and layout of the measurement}(b), then the $1\le W\le32$ antenna elements per AP are selected under the constraint $WN=M$. Thus, the system with the selected subarrays can be characterized as a distributed massive MIMO system when $M=64$ or $128$, a co-located MIMO system when $N=1$, and a general distributed MIMO system for other cases. At the user side, as mentioned in Section~II, `virtual’ users are selected from the positions measured along the robot's trajectory. Note that all the above selections are random, and Monto-Carlo simulations are performed to obtain statistical results for the analysis.     

First, the SVS performance of the measured channels is evaluated. For distributed massive MIMO channels with twelve closely located users, the cumulative distribution functions (CDFs) of the SVS measured in the different regions are shown in Fig.~\ref{Fig_SVS_DifferentRegion_12user}. The average SVS values for the LoS, mixed, and NLoS regions are 4.6 dB, 6 dB, and 4.1 dB, respectively. Due to the rich scattering environment and the subsequent `favorable' propagation, the averaged SVS of the NLoS region is the smallest, indicating better user channel orthogonality compared to the others. Compared to the LoS region, a larger averaged SVS is observed in the mixed region. One potential reason is that the dominant multipaths in the mixed region mainly come from the same direction, which contributes to a more similar multipath propagating behavior. This results in more correlated channels between closely located users, indicating poorer user orthogonality.   
\begin{figure}[tb]
	\centerline{\includegraphics[width=0.48\textwidth]{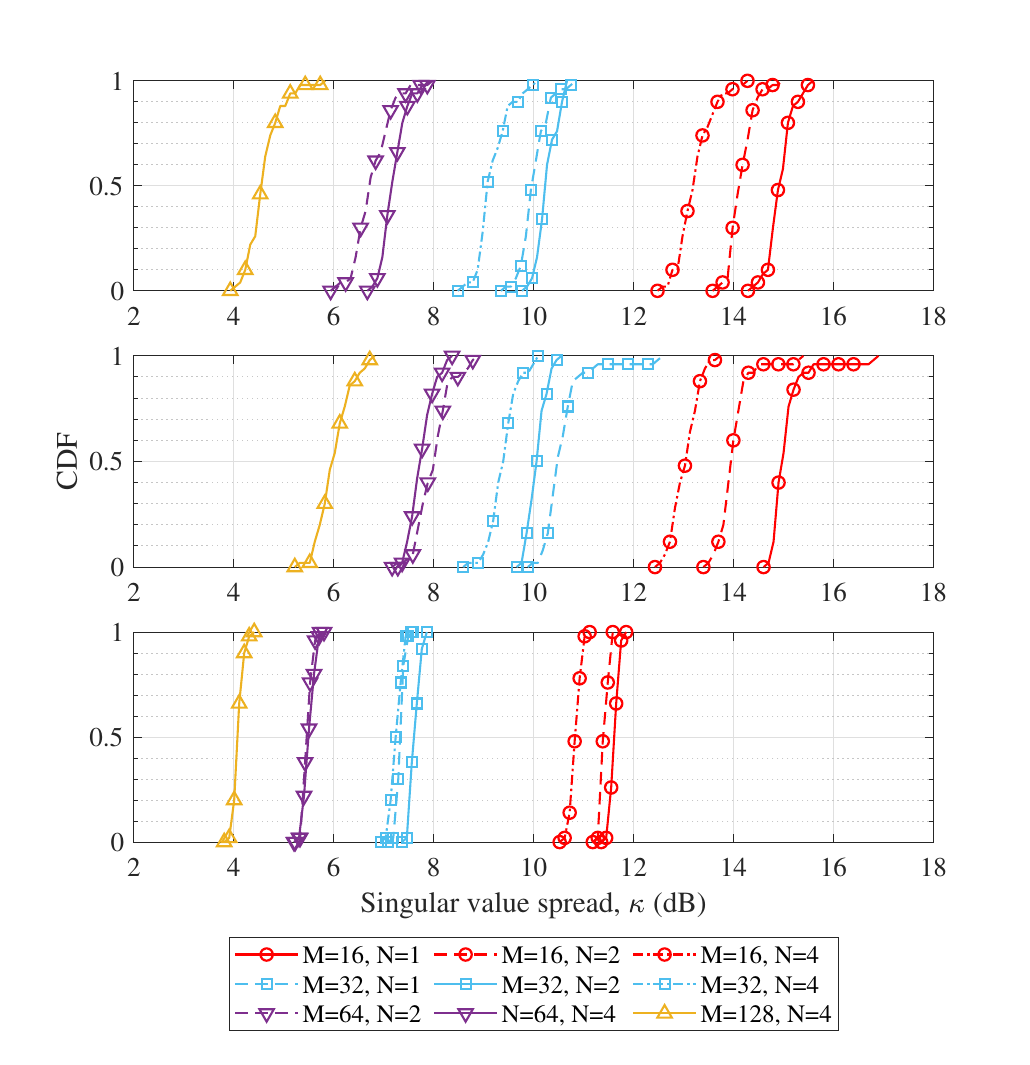}}
	\caption{CDFs of SVSs measured in LoS (top), mixed (middle), and NLoS (bottom) regions with different total number of antennas and antenna topologies ($K=12$).}
	\label{Fig_SVS_DifferentDistribution_DifferntRegion}
\end{figure}

Furthermore, the CDFs of the measured SVS with varying number of antennas and antenna topologies are illustrated in Fig.~\ref{Fig_SVS_DifferentDistribution_DifferntRegion}. It shows that for all measured regions, the SVS decrease as the total number of antennas increases from 16 to 32, 64, and 128. Furthermore, given the same number of antennas, increasing $N$ leads to a smaller SVS in the LoS region. This means that the user's channels become more orthogonal with a more distributed antenna topology. A similar phenomenon can be observed in the other two regions when the number of antennas is small. However, in the NLoS region, when a massive number of antennas is considered, further improvements of the user orthogonality cannot be obtained by deploying the antennas in a more distributed manner. This could be attributed to the inherently complex environment of NLoS propagation, which is supported by the fact that there is a larger azimuth angular spread of arrival ($64^\circ$) compared to other regions (e.g., $37^\circ$ in the LoS region), resulting in a limited improvement of the richness of the scatterers with a distributed antenna topology.    

\begin{figure}[tb]
	\centerline{\includegraphics[width=0.48\textwidth]{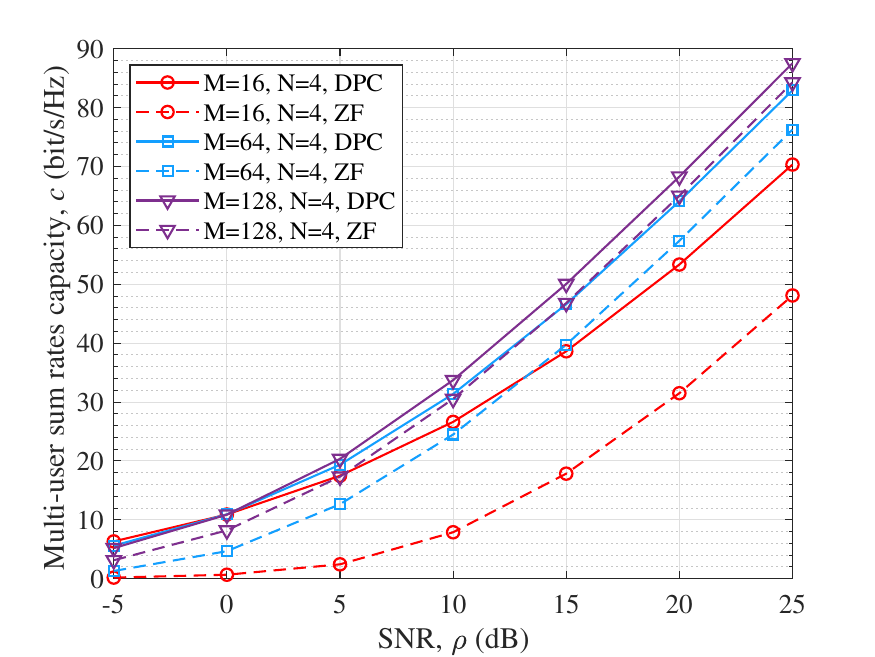}}
	\caption{Multi-user sum rates in the LoS region for the distributed MIMO system with different total number of antennas and precoding schemes ($K=12$).}
	\label{Fig_Capacity_DifferentPrecoder_6user_Region3}
\end{figure}

Comparisons of multi-user sum rates measured in the LoS region, utilizing the DPC and ZF precoding schemes, are presented in Fig.~\ref{Fig_Capacity_DifferentPrecoder_6user_Region3}. At the same SNR level, the differences in sum-rates capacities achieved by the DPC and ZF schemes tend to be insignificant as the total number of antennas is increasing. For instance, at $\rho=15$~dB, only $46\%$ of the DPC capacity (38.6 bit/s/Hz) is achieved by the ZF precoder scheme with $M=16$. This number increases to $85\%$ (of 49.9 bit/s/Hz) when $M=64$, and finally reaches $94\%$ (of 50.1 bit/s/Hz) when $M=128$. Similar trends are observed in channels measured in other regions, as summarized in TABLE~\ref{ZF Sum-Rates Measured in Different Regions}. The results indicate that, similar to the co-located massive MIMO system findings reported in~\cite{J. Flordelis2018}, the ZF linear precoder can achieve a large fraction of the DPC performance, but with lower complexity, in distributed massive MIMO systems.
\begin{table*}
\centering
\caption{Sum-rates measured in different regions with different precoder ($\rho =15$ dB, $K=12$, $N=4$).}
\label{ZF Sum-Rates Measured in Different Regions}
\begin{tabular}{cccccccccc}
\toprule
& \multicolumn{3}{c}{LoS region}& \multicolumn{3}{c}{Mixed region}& \multicolumn{3}{c}{NLoS region}\\ 
\midrule
Total number of antennas $M$ & 16& 64 & 128         & 16& 64 & 128         & 16& 64 &128         \\ 
\midrule
  DPC sum-rates (bit/s/Hz) & 38.6& 49.9& 50.1& 38.8& 46.8& 48.0& 42.6& 49.4&50.5\\ 
ZF sum-rates (bit/s/Hz)         & 17.9& 42.8&  46.8& 18.8& 38.8& 42.9& 26.8& 45.4&48.2\\
Fraction of DPC capacity achieved          & 46.2\%& 85.8\%&  93.5\%& 48.4\%& 82.9\%& 89.4\%& 63\%& 92\%&95.4\%\\
\bottomrule
\end{tabular}	
\end{table*}
\begin{figure}[tb]
	\centerline{\includegraphics[width=0.48\textwidth]{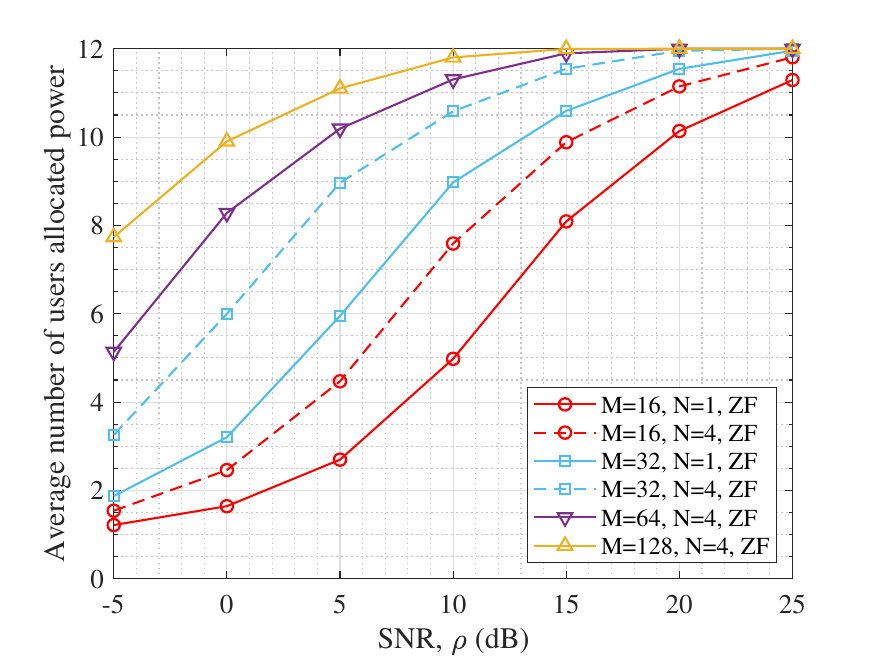}}
	\caption{Average number of users allocated power in the LoS region with the ZF precoder ($K=12$).}
	\label{Fig_ActiveUserNum_ZF_Region3}
\end{figure}

In a MU-MIMO system, it is recognized that to maximize the sum-rate capacity using water-filling algorithms, the AP/BS tends to allocate higher available data rates to those users experiencing advantageous channel conditions. For users with weak signal strengths, minimal or even no capacity is allocated. This leads to large imbalances in resource allocation, resulting in insufficient user fairness and exploitation of spatial multiplexing. To investigate the user fairness of the systems, the average number of users allocated power in the LoS region is given in Fig.~\ref{Fig_ActiveUserNum_ZF_Region3}, using the ZF precoding scheme. It reveals that systems equipped with a massive number of antennas can simultaneously schedule more users. For example, with $M=128$, all users can be allocated power at a moderate SNR value (e.g., $\rho =15$ dB), achieving non-zero communication for all users. In contrast, only ten or fewer users are scheduled in a distributed MIMO system equipped with only 16~antennas. Moreover, with the same number of antennas and SNR, a more distributed antenna topology demonstrates improved user fairness. Specifically, with $\rho =0$~dB, approximately three more users are allocated power, meaning non-zero communication, in a distributed MIMO system compared to a co-located MIMO system with the same number of antennas, ie $M=32$. 

\begin{figure}[tb]
	\centerline{\includegraphics[width=0.48\textwidth]{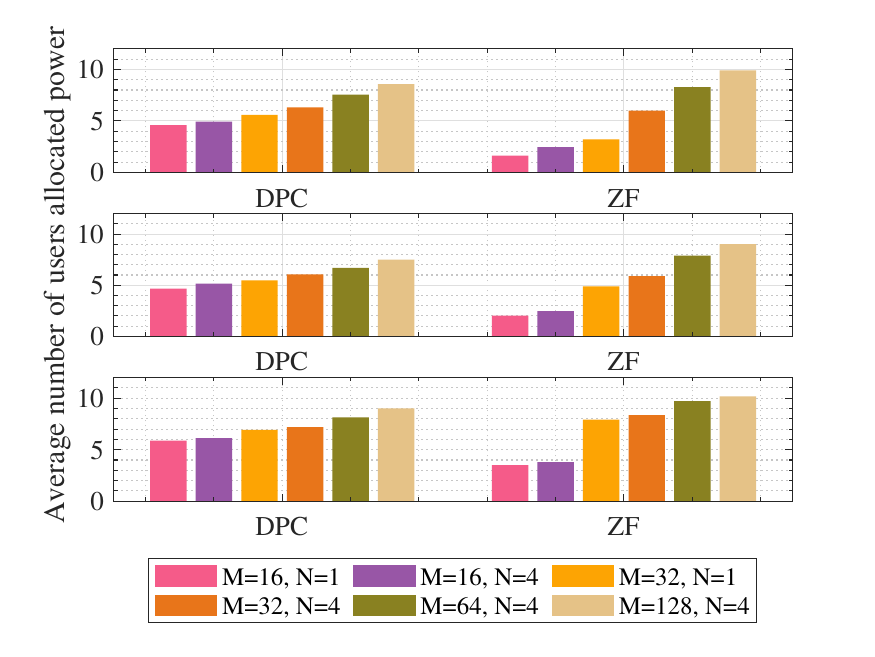}}
	\caption{Comparisons of the average number of users allocated power with the DPC and ZF precoders in LoS (top), mixed (middle), and NLoS (bottom) regions ($\rho =0$ dB, $K=12$).}
	\label{Fig_ActiveUserNum_SNR0dB_BarFigure}
\end{figure}

Comparisons of the average number of users allocated power with different precoding schemes in all regions are illustrated in Fig.~\ref{Fig_ActiveUserNum_SNR0dB_BarFigure}. It further demonstrates the superior spatial multiplexing of users in distributed MIMO systems compared to co-located MIMO systems with the same number of antennas. This is most evident with the ZF precoder in the LoS region. However, for the DPC scheme and in NLoS propagation, the improvement with a distributed antenna topology is limited. These results suggest that with a given number of antennas, it is preferable to deploy them more widely to achieve optimal user separability under LoS conditions. However, in the given NLoS condition, the differences in user separability between distributed and co-located antenna topologies are marginal. In addition, Fig.~\ref{Fig_ActiveUserNum_SNR0dB_BarFigure} also illustrates that, compared to the DPC scheme, a distributed massive MIMO system employing the ZF precoding scheme can serve more simultaneous users, exhibiting better spatial separability of the users. 
\section{Conclusions}
In this paper, distributed massive MIMO channel measurements have been performed in an indoor office. Based on the measured data, the spatial separation of closely-located users has been examined through multiple metrics, including singular value spread, DPC capacity, and ZF precoding sum-rate capacity. The system performance with different number of antennas and various antenna topologies has been compared in LoS, NLoS, and mixed scenarios. The results show that, compared to distributed MIMO with fewer antennas, distributed massive MIMO exhibits enhanced user orthogonality, larger sum-rate capacities, and better user fairness. Moreover, the ZF precoder can achieve a large fraction of the DPC capacity, but with lower complexity, in distributed massive MIMO systems. Furthermore, insights into antenna deployment strategies have been provided. Specifically, for LoS propagation, deploying antennas in a more distributed manner can enhance user separability considerably. However, for NLoS conditions in our case, the differences in user separability between distributed and co-located antenna topologies are marginal.


\section*{Acknowledgment}

This work was supported by the Swedish strategic research area ELLIIT and by REINDEER, which has received funding from the European Union’s Horizon 2020 research and innovation program under grant agreement No.
101013425.



%

\end{document}